\documentclass[
reprint,
amsmath, amssymb,
aps,
pra,
nolongbibliography
]{revtex4-2}
\usepackage{graphicx, xcolor}
\usepackage{soul}
\usepackage{hyperref}
\usepackage{comment} 

\providecommand*{\dd}[3][]{\frac{\mathrm{d}^{#1}#2}{\mathrm{d} #3^{#1}}}

\providecommand*{\qmbox}[1]{\quad \mbox{#1} \quad}

\renewcommand{\bm}[1]{\boldsymbol{\mathbf{#1}}}
\providecommand*{\qmbox}[1]{\quad \mbox{#1} \quad}
\newcommand{\G}{\mathcal{G}}
\providecommand{\F}{\mathcal{F}}
\providecommand{\R}{\mathcal{R}}
\newcommand{\bv}{\bm{v}}
\newcommand{\bx}{\bm{x}}
\newcommand{\br}{\bm{r}}
\newcommand{\bV}{\bm{V}}
\newcommand{\bX}{\bm{X}}
\newcommand{\bsigma}{\bm{\sigma}}
\newcommand{\avg}[1]{\left<#1\right>}

\newcommand{\intS}{\int_{\avg{S_p}}}
\newcommand{\intV}{\int_{\avg{V}}}

\begin{document}
\preprint{APS/123-QED}

\title{Time-averaged Dynamics of Compressible Particles in Oscillatory Gradient Flows}

\author{Xiaokang Zhang}
\author{Jake Minten}
\author{Bhargav Rallabandi}
\email{bhargav@engr.ucr.edu}
\affiliation{Department of Mechanical Engineering,
 University of California, Riverside, CA 92521
}

\date{\today}
\begin{abstract}
Acoustic fields effect steady transport of suspended particles by rectifying the inertia of primary oscillations. We develop a fully analytic theory that relates this steady particle motion to  incident oscillatory (acoustic) flow and the time-averaged force acting on the particle, systematically spanning the entire range between inviscid acoustofluidics and viscous particle hydrodynamics. By applying the Lorentz reciprocal theorem, we obtain a Fax\'{e}n-like relationship that includes nonlinear inertial forces, which depend on (i) the thickness of the oscillatory Stokes layer around the particle, and (ii) the density and compressibility contrast between the particle and the fluid.  The framework recovers secondary radiation forces for thin Stokes layers, and predicts a reversal of the motion when the thickness of the Stokes layer  is comparable to the particle size. We quantitatively validate the theory using numerical simulations of the timescale-separated hydrodynamics. 
  
\end{abstract}

\maketitle

The application of oscillatory fields is a powerful means to manipulate suspended particles and has recently been used in a wide range of applications, including microfluidic particle focusing and sorting  \cite{friend2011microscale, mutlu2018oscillatory}, cell patterning \cite{yang2022harmonic}, acoustic levitation \cite{lee2018collisional, andrade2020acoustic}, and the design of swimming microrobots \cite{klotsa2015propulsion}. An incident acoustic or otherwise oscillatory source excites an oscillatory flow around a suspended particle. The advective inertia of the primary oscillations drives a secondary flow that exerts a nonzero time-averaged force on the particle, leading to time-averaged motion of the particle along gradients of the incident field. For example, particles may accumulate at nodes or antinodes of an acoustic standing wave \cite{thomas2001dust}, be attracted to boundaries \cite{chen2016onset, mutlu2018oscillatory}, or may assemble into chains or clusters \cite{voth2002ordered, Klotsa_2009, lim2019cluster}. 

\begin{figure}[!ht]
\centering
\includegraphics[width=0.42\textwidth]{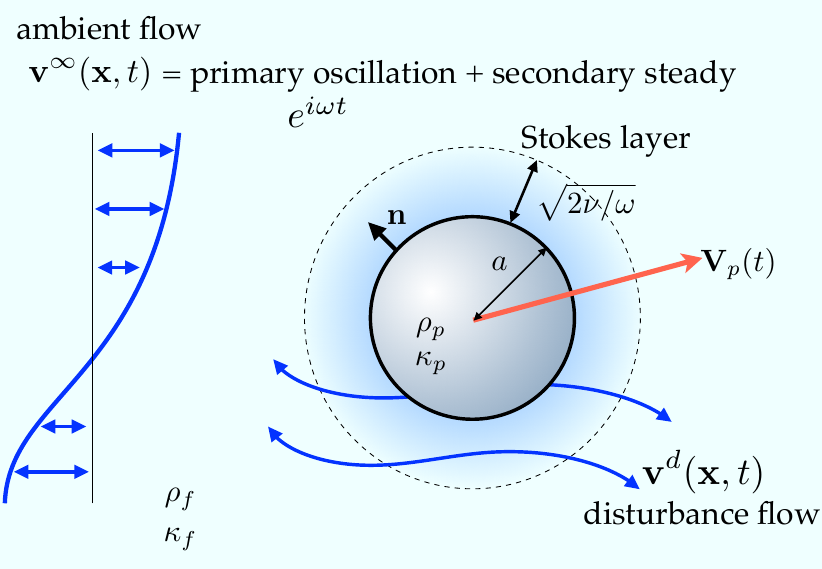}
\caption{An ambient flow of fluid (density $\rho_f$, compressibility $\kappa_f$) produces oscillations of a suspended particle (density $\rho_p$, compressibility $\kappa_p$). Advective nonlinearities drive a secondary time-averaged  motion of the particle.}
\label{fig.setup}
\end{figure}

The flow is controlled by the ratio $\delta = \sqrt{\frac{2 \nu}{\omega a^2 }}$ of a viscous Stokes layer thickness to the particle radius $a$ ($\nu$ is the kinematic viscosity of the fluid and $\omega$ is the angular frequency of oscillation); see Fig. \ref{fig.setup}.  In the inviscid acoustic limit ($\delta \ll 1$), the time-averaged particle dynamics are well understood through the theory of secondary radiation forces \cite{king1934acoustic, settnes2012forces}. An alternative (better suited for $\delta \gg 1$) uses the Gatignol--Maxey--Riley equation \cite{gat83,max83} (often with modifications \cite{cho13_inertial,thameem2017fast,agarwal2018inertial,agarwal2023densitycontrast}) but neglects compressibility effects important in acoustics. Most applications operate at intermediate $\delta$, where no simple analytic theory exists, and where the above approaches and direct hydrodynamic calculations \cite{doinikov1994acoustic,danilov2000mean} can yield contradictory predictions for the particle dynamics.

In this Letter, we develop  analytic theory and numerical simulations for the time-averaged motion of a spherical particle in an oscillatory flow, systematically accounting for inertial and viscous forces (arbitrary $\delta$) and compressibility effects. We start with a known ambient (or incident) fluid flow $\bv^{\infty}(\bx,t)$ that is defined in the absence of the particle (Fig. \ref{fig.setup}) and is characterized by a combination of an oscillatory primary component of characteristic speed $v$ and a slower 
secondary component with non-zero time-average. 
Such flows are common in nonlinear acoustics and in streaming flows driven by oscillating boundaries. The flow in the presence of the particle is $\bv(\bx, t) = \bv^{\infty}(\bx, t) + \bv^d(\bx, t)$, where $\bv^d$ is the disturbance (or scattered) flow produced by the particle. The fluid ($f$) and the particle ($p$) have equilibrium density $\rho_{f,p}$ and compressibility $\kappa_{f,p} = (\rho_{f,p} c_{f,p}^2)^{-1}$, where $c_{f,p}$ is the speed of sound in the medium. Scaling length with $a$, time with $\omega^{-1}$ and defining a dimensionless density field $\varrho(\bx ,t) = \rho(\bm{x},t)/\rho_f$, the flow is governed by
\begin{subequations} \label{NS}
\begin{align}
\frac{2}{\delta^2}\left(\frac{\partial\bv}{\partial{t}}+\varepsilon\bv\cdot\nabla\bv\right)=&\nabla\cdot\bsigma,\\
\frac{\partial\varrho}{\partial{t}}+\varepsilon\nabla\cdot(\varrho\bv)=&0\,.
\end{align}
\end{subequations}
Here, 
$\bsigma=-p\bm{I}+\left(\nabla\bv+\nabla\bv^\mathsf{T}\right){}$ is the stress tensor (scaled with $\mu v/a$, where $\mu = \nu \rho_f $) and $\varepsilon = v/(a \omega)$ is the dimensionless amplitude of oscillation. The particle translates [velocity $\bm{V}_p(t)$], 
and undergoes volume oscillations (with a surface velocity $V_n(t) \bm{n}$, $\bm{n}$ being the fluid-facing unit normal). On the particle surface $S_p(t)$,  the flow thus satisfies 
\begin{align} 
    \bv(\bm{x},t)=\bV_p(t)  + V_n(t) \bm{n}, \quad \bm{x} \in S_p(t),
\label{eqn.bc}
\end{align}
Rotation of the particle does not contribute to the force due to symmetry \cite{danilov2000mean}, so we neglect it here.  

We seek to relate the time-averaged motion of the particle to the (known) ambient flow and time-averaged forces acting on the particle for arbitrary $\delta$ and small oscillation amplitude $\varepsilon \ll 1$ \footnote{A frame-invariance argument finds that the amplitude of particle oscillation \emph{relative} to the ambient flow must be small, i.e. $|\varepsilon \R| \ll 1$.}. We invoke a perturbation solution with $(\bv, \bsigma) \sim (\bv_1, \bsigma_1) + \varepsilon(\bv_2, \bsigma_2)$ and $\varrho \sim 1 + \varepsilon \varrho_1$.  
Primary components (subscript 1) are strictly oscillatory, whereas secondary components (subscript 2) additionally involve steady components, which are of interest. Separating orders of $\varepsilon$ in \eqref{NS} leads to
\begin{subequations}\label{SeparatedGE}
  \begin{align} 
   & \frac{2}{\delta^2}\frac{\partial\bv_1}{\partial t}=\nabla\cdot\bsigma_1,\quad\frac{\partial\varrho_1}{\partial t}+\nabla\cdot\bv_1=0. \label{eqn.1stGE} \\
   &{\nabla}\cdot\avg{{\bsigma}_2-\frac{2}{\delta^2}{\bv}_1{\bv}_1}=\bm{0}
     ,\quad \nabla\cdot\avg{\bv_2 + \varrho_1 \bv_1} = 0, \label{eqn.2ndGE}
\end{align}  
\end{subequations}
where angle brackets define a time-average over an oscillation according to $\left<g\right>(\bm{x}) = (2 \pi)^{-1} \int_{t}^{t + 2\pi} g(\bm{x}, t)$ and isolate steady flow features.  As is typical in acoustics, the primary flow is weakly compressible, with pressure and density oscillations being related by $p_1= \varrho_1 c_f^2/(\nu\omega)$. The inertia of the secondary flow is typically small \footnote{This corresponds to the condition $\varepsilon \ll \delta$.} and has been neglected in \eqref{eqn.2ndGE}.

Similarly expanding the particle kinematics into primary and secondary contributions, projecting \eqref{eqn.bc} onto the mean particle surface $\left<S_p \right>$, and separating powers of $\varepsilon$ yields effective boundary conditions (details in Supplemental Material \cite{SIFoot})
\begin{subequations} \label{SeparatedBCs}
   \begin{align} 
    \bv_1&=\bV_{p1} + \bV_{n1}\bm{n}  \qmbox{for} \bm{x} \in \left<S_p\right>, \label{eqn.bcv1}\\
    \bv_2&=\bV_{p2}-\left<\int\bv_1dt\cdot\nabla\bv_1\right> \qmbox{for} \bm{x} \in \left<S_p\right> \label{eqn.bcv2}.
\end{align} 
\end{subequations} 

We first solve for the primary (oscillatory) flow around the particle \cite{oppenheimer2016_hot,danilov2000mean} by making the ansatz that they are of the form $\text{Re}\left[g(\bm{x}) e^{i t}\right]$ for generally complex $g(\bm{x})$. Spatial variations of the ambient flow occur on length scales much larger than the particle (either the wavelength of sound $L_c = 2\pi c_f/\omega$ or a geometric scale $L_g$). Defining a time-averaged (i.e. inertial) frame $\bm{r} = \bm{x} - \avg{\bm{X}_p}$ centered at the \emph{time-averaged particle center} $\avg{\bm{X}_p}$, we expand the primary ambient flow as 
\begin{align} \label{Vinf}
\bv_1^\infty(\bm{x}, t) \sim \bV_1^\infty(t)+ \bm{E}_1^{\infty}(t)\cdot\br+ \frac{1}{3}{\Delta}_1^{\infty}(t)\br + \dots 
\end{align}
in terms of the velocity $\bm{V}_1^{\infty}$, the deviatoric rate of strain (i.e. extension rate) $\bm{E}_1^{\infty}$ and the velocity divergence $\Delta_1^{\infty}$ of the ambient flow, all evaluated at  $\bx = \avg{\bX_p}$. Note that these flow properties are complex phasors, and it will be understood that only the real part of any complex equality is physically meaningful. The vorticity of the ambient flow does not contribute to forces due to symmetry and has been neglected in \eqref{Vinf}. Solving \eqref{eqn.1stGE}, \eqref{eqn.bcv1} for $a \ll L_c$ yields the primary disturbance flow 
\begin{align}
   \!\! \bv_1^d= \bm{D}\cdot(\bV_{p1} - \bV_1^{\infty}) + \bm{\mathcal{Q}}:\bm{E}_1^\infty +  \bm{m} \left(\! V_{n1} -\frac{\Delta^{\infty}_1}{3}\! \right)\!,
\label{eqn.dist1st}
\end{align}
where $\bm{m}(\bm{r})$, $\bm{D}(\bm{r},\delta)$ and $\bm{\mathcal{Q}}(\bm{r},\delta)$ are well-known monopole (rank-1), dipole (rank-2) and quadrupole (rank-3) tensor solutions; see \cite{SIFoot}. 

\begin{figure}[!ht]
\centering
\includegraphics[width=0.27\textwidth]{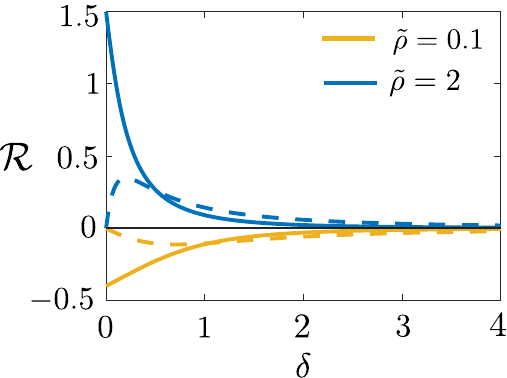}
\caption{Relative particle mobility, showing real (solid; in-phase) and imaginary (dashed; out-of-phase) parts.}
\label{Fig.R}
\end{figure}

The primary oscillatory flow $\bv_1 = \bv_1^{\infty} + \bv_1^d$ is now known (as is the primary stress $\bsigma_1$), up to the oscillatory particle kinematics $\bm{V}_{p1}$ and $V_{n1}$.  To this end, we invoke conservation of the particle momentum (projected on $e^{i t}$ modes), $\frac{4}{3} \pi a^3 \rho_p  \dd{\bV_{p1}}{t} = \intS \bm{n} \cdot \bsigma_1 dS$. This establishes the oscillatory velocity of the particle relative to that of the ambient flow according to \cite{settnes2012forces}
\begin{subequations}
\begin{align}
    \bm{V}_{p1} - \bm{V}_{1}^{\infty} &= \R \bm{V}_{1}^{\infty}, \qmbox{where}\\
    \mathcal{R}(\lambda, \tilde{\rho})&=-\frac{2\lambda^2(\Tilde{\rho} - 1)}{\lambda^2 (2 \Tilde{\rho}+ 1) + 9 \lambda + 9} 
\end{align}
\end{subequations}
is a relative particle mobility (see Fig. \ref{Fig.R}), $\tilde{\rho} = \rho_p/\rho_f$ is the density ratio, and $\lambda = (1 + i)/\delta$ is a complex reciprocal Stokes layer thickness. Real and imaginary parts of $\R$,  respectively, quantify in-phase and out-of-phase oscillations of the particle relative to the fluid. Similarly, equilibrium of normal stresses on the particle surface determines $V_{n1} = \frac{1}{3}\left(\tilde{\kappa} - 1 \right) \Delta_1^{\infty}$, where $\tilde{\kappa} = \kappa_p/\kappa_f$ is the compressibility ratio \cite{settnes2012forces}. 

Having fully determined the primary flow $\bm{v}_1$, we turn to the time-averaged particle motion.  Doinikov \cite{doinikov1994acoustic} showed that the average force exerted by the fluid on the particle (in units of $\epsilon \mu a v$) is
\begin{align} 
    \left<\bm{ F}\right>=\intS\bm{ n}\cdot\left<\bsigma_2-\frac{2}{\delta^2} \bv_1\bv_1\right> \,dS.
\label{eqn.doi}
\end{align}
To make an analytic prediction for $\left<\bm{ F}\right>$ without calculating $\bm{\sigma}_2$ in detail (which requires a solution to the secondary flow), we reformulate \eqref{eqn.doi} using the Lorentz reciprocal theorem \cite{masoud2019reciprocal}. We introduce, as an auxiliary flow, the steady, incompressible, Stokes flow [velocity $\hat{\bv}(\bx)$, stress $\hat{\bsigma}(\bx)$] produced by a sphere translating with velocity $\hat{\bV}$ through quiescent fluid. The rate of strain of this auxiliary flow is $\hat{\bm{E}}(\bm{x}) = \bm{\mathcal{E}}(\bm{x}) \cdot \hat{\bV}$ and the auxiliary traction on the particle surface is $\bm{n}\cdot \hat{\bsigma}|_{<S_p>} = \bm{T}(\bm{x}) \cdot \hat{\bV}$, where the tensors $\bm{\mathcal{E}}$ (rank 3) and $\bm{T}$ (rank 2) are well-known (e.g. \cite{oppenheimer2016_hot,kim1993microhydrodynamics,SIFoot}). Starting with \eqref{eqn.2ndGE}, we construct the symmetry relation $\nabla \cdot \left<\bsigma_2 - \frac{2}{\delta^2} \bv_1 \bv_1 \right> \cdot \hat{\bv} = \nabla \cdot \hat{\bsigma} \cdot \bv_2$ and integrate over the fluid volume to recast \eqref{eqn.doi} as (see SI)
\begin{align} \label{FRT}
\left<\bm{F}\right>&=\intS\left(\bV_{p2}-\bv_{2L}^{\infty} - \left<\int\bv_1dt\cdot\nabla\bv_1\right>^d\right)\cdot\bm{T}dS \nonumber \\
&\quad +\intV \frac{2}{\delta^2}\left<\bv_1\bv_1\right>^d:\bm{\mathcal{E}}  dV,
\end{align}  
where $\left<V\right>$ represents the volume surrounding the time-averaged particle surface. Above, we have introduced the (known) ambient time-averaged Lagrangian ``streaming'' velocity $\bv_{2L}^{\infty}(\bm{x})=\avg{\bv^{\infty} + \int\bv_1^{\infty}dt\cdot\nabla\bv_1^{\infty}}$, which represents the average velocity of a \emph{material} fluid element in the absence of the particle \cite{riley2001steady}. Using standard averaging rules for products of complex oscillating quantities \cite{longuet1998viscous}, we find that the time-averaged force on the particle (reverting to dimensional variables) is 
\begin{align} \label{FResult}
\left<\bm{F}\right>&\stackrel{\text{Real}}{=}-6\pi\mu a\left\{\bm V_{p2}-\bv_{2L}^{\infty}-\frac{a^2}{6}\nabla^2\bv_{2L}^{\infty}\right\}\bigg|_{\bx = \avg{\bm{X}_p}} \nonumber  \\
&\qquad + m_f  \left(\bV^{\infty}_1\right)^*\cdot\bm{E}_1^\infty \; \F_E( \lambda, \tilde{\rho}) \nonumber \\  &\qquad +m_f \left(\bV^{\infty}_1\right)^*{\Delta}_1^\infty \; \F_\Delta( \lambda, \tilde{\rho}, \tilde{\kappa}),
\end{align}
where $m_f = \frac{4}{3} \pi a^3 \rho_f$, and $\F_E$ and $\F_\Delta$ are complex coefficients that we discuss in detail later. The asterisk denotes a complex conjugate, and only the real part of \eqref{FResult} is physically relevant \footnote{An equivalent formulation without complex variables is in \cite{SIFoot}}.  

\begin{figure}[t!]
\centering
\includegraphics[width=0.48\textwidth]{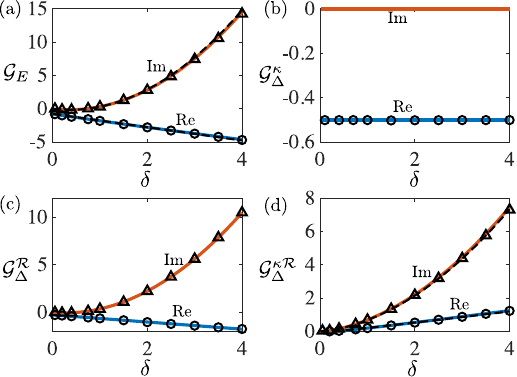}
\caption{Coefficients (a) $\G_E$, (b) $\G_{\Delta}^{\kappa}$, (c) $\G_{\Delta}^{\R}$ and (d) $\G_{\Delta}^{\kappa \R}$ as obtained from the theory (curves) and from numerical solutions (symbols), showing real and imaginary parts. Approximations (\ref{Gcoeffs}a,d) are indicated by dashed curves in panels (a,d). } 
\label{Fig.AllGCoeffs}
\end{figure}

The first term of \eqref{FResult} is a Stokes drag with a Fax\'{e}n correction for a non-inertial particle moving through the \emph{Lagrangian} ambient streaming field $\bm{v}_{2L}^{\infty}$. The second and third terms are inertial forces that depend on a quadratic combinations of the ambient oscillatory velocity and rate of strain. The associated complex coefficients $\F_{E, \Delta}$ determine the strengths of these forces and account for both in-phase and out-of-phase oscillations.  For example, the real part of $\F_{E, \Delta}$ quantifies the force resulting from an ambient flow velocity that oscillates in phase with the ambient strain-rate (e.g. a standing acoustic wave), whereas $\F_{E, \Delta}$ characterizes forces due to  $90^{\circ}$ out-of-phase oscillations (e.g. a traveling wave). We find that these coefficients admit the exact decomposition 
\begin{subequations} 
\begin{align}
    \F_E &= \R^* \G_E, \\
    \F_\Delta &= \left(\tilde{\kappa} - 1\right) \G_{\Delta}^{\kappa} +  \R^* \G_{\Delta}^{\R}+ (\tilde{\kappa} - 1) \R^* \G_{\Delta}^{\kappa \R},
\end{align}
\end{subequations}
into terms that depend on the density contrast $(\tilde{\rho} -1)$ (through $\R^*$), the compressibility contrast $(\tilde{\kappa} -1)$ or a product of the two. The associated complex coefficients $\G_A^B(\lambda)$ are purely hydrodynamic quantities (independent of particle properties) that arise from the spatial structure of the primary flow. They are obtained by analytic evaluation of the integrals in \eqref{FRT}  (using \emph{Mathematica}); see solid curves in Fig. \ref{Fig.AllGCoeffs}. Simple expressions for these coefficients are given by
\begin{subequations} \label{Gcoeffs}
\begin{align}
    \G_E(\lambda) &\simeq   -\frac{3 \lambda ^2 + \frac{4}{5}(1+9 i)  \lambda +9}{4 \lambda ^2} \label{GcoeffsE},\\
    \G_{\Delta}^{\kappa}(\lambda) &= -\frac{1}{2} \label{GcoeffsDelkappa},\\
    \G_{\Delta}^{\R}(\lambda) &= -\frac{\lambda ^2+3 i \lambda +6}{4 \lambda ^2} \label{GcoeffsDelR},\\
    \G_{\Delta}^{\kappa \R}(\lambda) &\simeq -\frac{15}{4 \lambda^2} \frac{ (9 \lambda +8 i)}{(9 \lambda +40 i)} \label{GcoeffsDelmix}.
\end{align}
\end{subequations}
Note that (\ref{Gcoeffs}b,c) are exact, while the approximations (\ref{Gcoeffs}a,d) are accurate to within $3\%$ of the exact results \cite{SIFoot} and are asymptotic at leading order for both small and large $\lambda$; see dashed curves in Fig. \ref{Fig.AllGCoeffs}(a,d). 

The relations \eqref{FResult}--\eqref{Gcoeffs} describe in full the time-averaged motion of a particle suspended in an oscillatory gradient flow and form the main theoretical result of this Letter. In the inviscid limit of $\delta \to 0$ ($\lambda \to \infty$), the present formulation fully recovers the theory of secondary radiation forces \cite{king1934acoustic, settnes2012forces}, while the viscous limit $\delta \to \infty$ recovers Stokesian hydrodynamics; cf. \cite{nadal2016small}. We note that time-averaged inertial force contributions due to flow curvature \cite{agarwal2021unrecognized} can simply be added to the right hand side of \eqref{FResult}. Furthermore, because the secondary flow is quasi-steady and inertialess, the sum of the time-averaged hydrodynamic force $\avg{\bm{F}}$ and external non-hydrodynamic forces (e.g., particle's buoyant weight) is zero. This condition thus determines the time-averaged velocity $\bm{V}_{p2}$ of a freely suspended particle. 

\begin{figure*}[!t]
\centering
\includegraphics[width=0.98\textwidth]{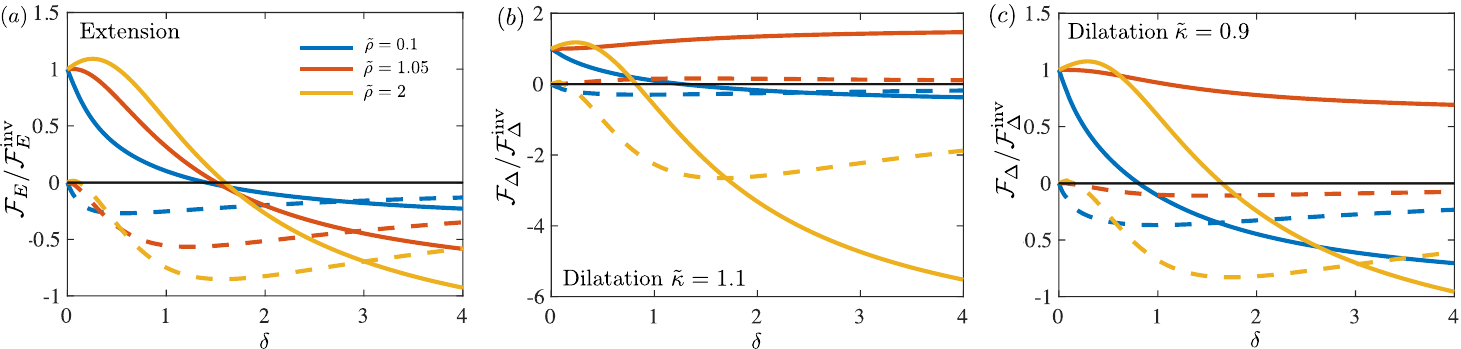}
\caption{Real (solid) and imaginary (dashed) parts of (a) $\mathcal{F}_E$ and (b,c) $\mathcal{F}_{\Delta}$ (normalized by their inviscid limits) for different density and compressibility ratios. The change in the signs of the real (imaginary) parts of the coefficients indicates a reversal in the direction of the force due to straining components that are in phase (out of phase) with the fluid velocity.} 
\label{Fig.FCoeffs}
\end{figure*}

We now discuss the behavior of the inertial force contributions of \eqref{FResult} in detail. The coefficient $\F_E = \R^* \G_E$ associated with extensional flow is nonzero only for density-mismatched particles, and approaches real-valued constants in both the inviscid [$\delta \ll 1$, $\F_E \to \F_E^{\rm inv} = \frac{3 (\tilde{\rho}-1)}{2(2 \tilde{\rho} + 1)}$], and the viscous [$\delta \gg 1$ , $\F_E \to \F_E^{\rm visc} = -\frac{(\tilde{\rho}-1)}{2}$] limits, in agreement with \cite{agarwal2023densitycontrast}. Thus, only velocities oscillating in phase with the extension rate lead to time-averaged forces in either limit. Notably, the inviscid and viscous limits are of opposite sign, indicating a reversal of the corresponding inertial force contribution with $\delta$ (Fig. \ref{Fig.FCoeffs}a). This reversal occurs when $\delta \approx 1.5$ and increases weakly with density ratio (Fig. \ref{Fig.FCoeffs}a; see also \cite{SIFoot}). The imaginary part of $\F_E$ vanishes in both limits, and achieves a maximum at intermediate $\delta/a$. Both real (in-phase) and imaginary (out-of-phase) parts of $\F_E$ are comparable for the $O(1)$ values of $\delta$ typical of applications and are both likely to be important in  practical oscillatory flows.  

The contribution of dilatation is somewhat more complicated as it depends on all three physical parameters ($\delta$, $\tilde{\rho}$ and $\tilde{\kappa}$); see \eqref{Gcoeffs}. As with $\F_E$, the coefficient $\F_{\Delta}$ [Fig. \ref{Fig.FCoeffs}(b,c)] asymptotes to real-valued constants in both the inviscid [$\F_{\Delta}^{\rm inv} = -\frac{\tilde{\kappa}-1}{2} + \frac{(\tilde{\rho}-1)}{2(2 \tilde{\rho} + 1)}$] and viscous [$\F_{\Delta}^{\rm visc} = -\frac{1}{6} (\tilde{\kappa}-1) (\tilde{\rho} + 2)-\frac{1}{3} (\tilde{\rho}-1)$] limits. We plot the coefficient $\F_{\Delta}$  (after normalizing by $\F_{\Delta}^{\rm inv}$) for different density and two compressibility ratios in Fig. \ref{Fig.FCoeffs}(b,c). While $\F_{\Delta}$ may change sign with $\delta$, this feature is not universal and only occurs for a limited range of density and compressibility ratio. 

\begin{figure}[!ht]
\centering
\includegraphics[width=0.49\textwidth]{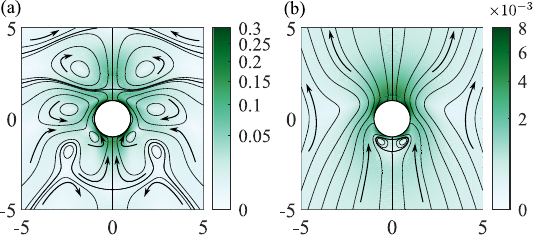}
\caption{Time-averaged disturbance flow around a polystyrene sphere in water ($\tilde{\rho} = 1.05$, $\tilde{\kappa} = 0.34$) with $\delta=1$ and $\bm{V}_{p2} = \bm{v}_{2L}^{\infty} = \bm{0}$. The flow is axisymmetric about the polar axis $\bm{p}$. Colors indicate flow speed. Primary ambient flows correspond to (a) $\bm{E}_1^\infty=\frac{1}{2}(-\bm{I}+3\bm{p} \bm{p})\, e^{it}$, ${\Delta}_1^\infty=0$, and (b) $\bm{E}_1^\infty=\bm{0}$, ${\Delta}_1^\infty=e^{it}$;  $\bV_1^\infty=\bm{p}\,e^{it}$ in both cases. }
\label{Fig.PS_in_wa}
\end{figure}

Finally, we verify the predictions of our theory using numerical solutions of the detailed flow in an axisymmetric setting under the small-amplitude perturbation scheme \eqref{SeparatedGE}--\eqref{SeparatedBCs}. We use the analytical formulation of the oscillatory flow (described above) and numerically solve for the secondary flow in detail, holding the particle stationary on average, and with no ambient Lagrangian streaming ($\bm{V}_{p2} = \bm{v}_{2L}^{\infty} = \bm{0}$). We then use \eqref{eqn.doi} to calculate $\avg{\bm{F}}$, which under the above setup isolates the inertial contributions of \eqref{FResult}. Furthermore, constructing oscillatory flows with pairs of flow modes lets us identify the computed force with a single term in \eqref{FResult} (e.g., a simulation with $\Delta_1^{\infty}=0$ and nonzero $\bm{V}_1^{\infty}$ and $\bm{E}_1^{\infty}$ identifies $\G_E$). The $\G$ coefficients thus computed are in excellent agreement (typically to within 4\%) with the exact results; see symbols in Fig. \ref{Fig.AllGCoeffs}. 

Figure \ref{Fig.PS_in_wa} shows example streamlines of the numerically computed secondary flow associated for a polystyrene bead in water (see also \cite{SIFoot}). Though quite complex near the sphere, the secondary flow exhibits the  $r^{-1}$ far-field velocity decay characteristic of Stokes flow driven by a point force. Interestingly, the associated point force is distinct from $\avg{\bm{F}}$, as a part of the secondary stress (viz., the radiation pressure) is in \emph{hydrostatic} balance with a part of the Reynolds stress $\avg{\frac{2}{\delta^2} \bv_1 \bv_1}$ and does not engender a secondary flow. A detailed exposition of these features is left to future work. 

Whether extension or dilatation ultimately dominates the time-averaged particle dynamics depend on the details of the ambient flow as well as the physical properties of the system. When the ambient flow surrounds (or is generated by) a geometric feature of size $L_g \ll L_c$ (e.g. in microstreaming flows), the extension rate $\propto v/L_g$ is much greater than the dilatation rate $\propto v/L_c$. In this case, the extensional component of inertial force dominates the dilatational one, provided that the properties of the particle do not contrast too strongly with that of the fluid. In the same geometric situation, however, both contributions may be important for large density or compressibility contrasts (e.g. a surfactant-coated gas bubble in water). By contrast, acoustofluidic and acoustic levitation setups typically use $L_g \simeq L_c$, so both extensional and dilatational contributions are equally important at the outset. The fully analytic theory developed here encompasses all of these situations over the entire range of $\delta$, $\tilde{\rho}$ and $\tilde{\kappa}$, and is thus a powerful quantitative tool to understand the dynamics of suspended objects in a wide range of acoustic and oscillatory flow systems. \\


We are grateful to S. Agarwal,  M. Gazzola and S. Hilgenfeldt for stimulating discussions, and thank the National Science Foundation for support through grant CBET-2143943. 


%

\end{document}